\documentclass{article}
\usepackage{spconf,amsmath,graphicx}
\usepackage{multirow}
\usepackage{amssymb}
\usepackage{pifont}
\usepackage{xcolor}
\usepackage{url}
\usepackage{svg}

\usepackage{enumitem}
\setlist{nosep, leftmargin=14pt}

\usepackage{mwe} 

\title{Assessing the Efficacy of Invisible Watermarks in AI-Generated Medical Images}
\name{Xiaodan Xing$^{\star}$ \qquad Huiyu Zhou$^{\dagger}$ \qquad Yingying Fang$^{\ddagger}$ \qquad Guang Yang$^{\star \ddagger \mathsection}$}
\address{$^{\star}$ Department of Bioengineering, Imperial College London, London, UK \\
         $^{\dagger}$  Computing and Mathematic Sciences, University of Leicester, Leicester, UK\\
         $^{\ddagger}$ National Heart and Lung Institute, Imperial College London, London, UK\\
         $^{\mathsection}$ Cardiovascular Research Centre, Royal Brompton Hospital, London, UK\\}
\begin{document}
%
\maketitle
\begin{abstract}
AI-generated medical images are gaining growing popularity due to their potential to address the data scarcity challenge in the real world. However, the issue of accurate identification of these synthetic images, particularly when they exhibit remarkable realism with their real copies, remains a concern. To mitigate this challenge, image generators such as DALLE and Imagen, have integrated digital watermarks aimed at facilitating the discernment of synthetic images' authenticity. These watermarks are embedded within the image pixels and are invisible to the human eye while remains their detectability. Nevertheless, a comprehensive investigation into the potential impact of these invisible watermarks on the utility of synthetic medical images has been lacking. In this study, we propose the incorporation of invisible watermarks into synthetic medical images and seek to evaluate their efficacy in the context of downstream classification tasks. Our goal is to pave the way for discussions on the viability of such watermarks in boosting the detectability of synthetic medical images, fortifying ethical standards, and safeguarding against data pollution and potential scams. 
\end{abstract}
%


\section{Introduction}
\label{sec:introduction}
The use of AI-generated medical images is increasing, as they offer a potential solution to the problem of limited data in real-world medical situations. Studies \cite{chawla2002smote,coyner2022synthetic} have shown that synthetic medical images can improve the accuracy of various medical tasks. However, there is a challenge in distinguishing between these synthetic images and real medical images because they can look very similar. This lack of distinction can lead to fraudulent use of realistic synthetic images if they are not properly labeled as synthetic \cite{qi2020emerging}. In addition, it is also demonstrated in the literature that exploiting generative images may introduce noise or impurities to large scale downstream tasks \cite{hataya2023will}.
\begin{figure}
    \centering
    \includegraphics[width=\linewidth]{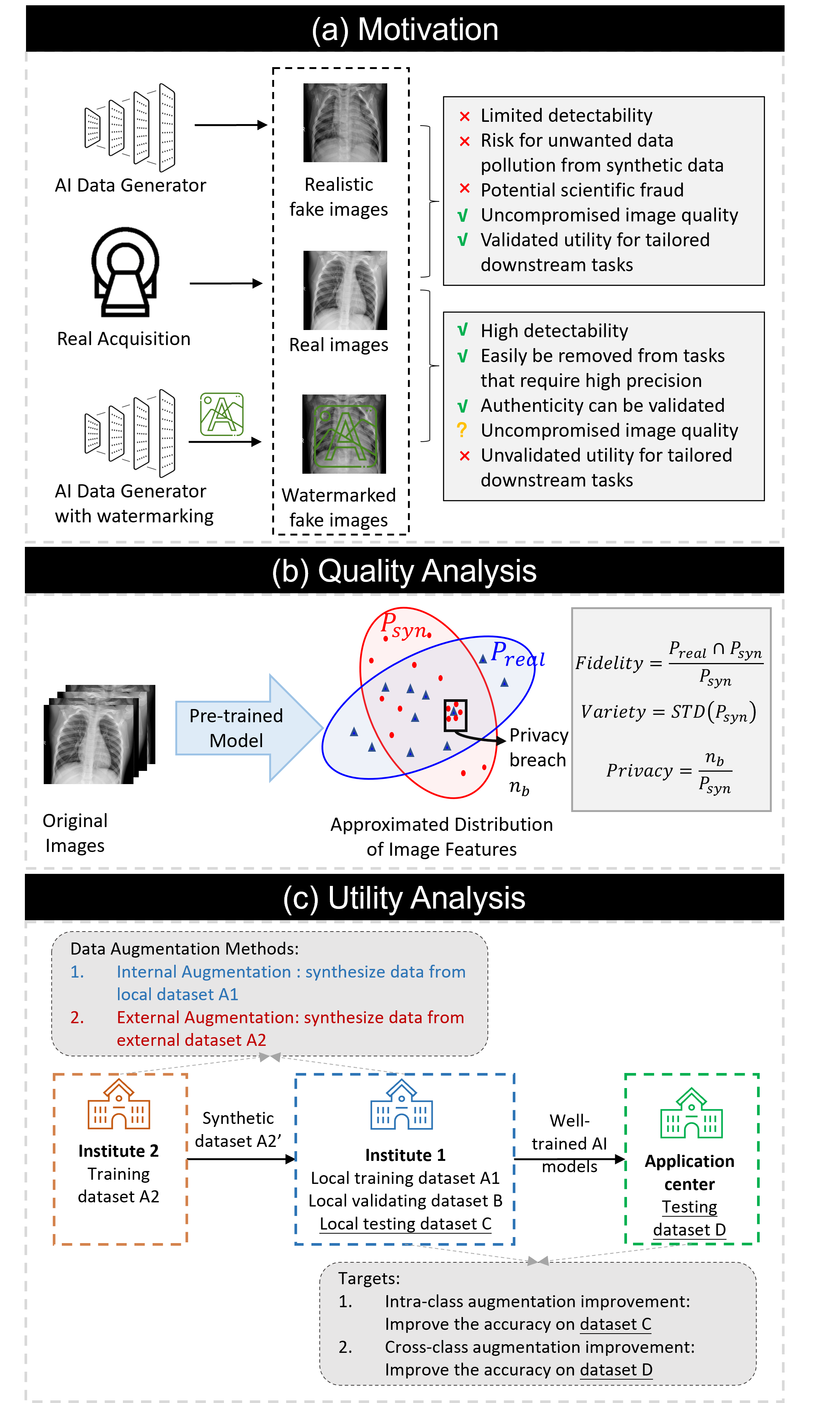}
    \caption{A graphical summary of our paper including our motivation of evaluating the quality and utility of water-marked synthetic images (a), and the method we used to analyze quality (b) and utility (c).}
    \label{fig:utility}
\end{figure}
To address these ethical considerations, researchers have explored adding identifiers, such as invisible watermarks \cite{fang2016reversible}, to synthetic images to help users distinguish between real and synthetic ones. These watermarking techniques are designed to be imperceptible to the human eye. SynthID \cite{synthid}, used in Imagen \cite{saharia2022photorealistic}, is a technique that effectively marks synthetic images with invisible watermarks, making them identifiable to users. In addition, instead of adding watermarks to synthesized images, researchers are also working on adding invisible water marks directly to generative models \cite{fernandez2023stable}. 

These invisible watermarking techniques are advocated for not only improving the detectability of synthetic images, but also maintaining unnoticeable to human observers. However, in the context of synthetic medical images, it is not only the visual appearance that downstream users require. In medical image synthesis, the utility of synthetic images for downstream tasks holds paramount importance. With improved authenticity and detectability, we do not want to compromise the primary utility of these synthesized medical images in real-world applications. However, unfortunately, there is currently scarce of literature exploring the impact of these invisible watermarks on the utility aspect of synthetic medical images. 

This paper addresses this gap by adding invisible watermarks to synthetic medical images from a large X-ray dataset generated by two popular synthesis algorithms. We evaluate whether adding watermarks affects the data augmentation ability of these images in multiple scenarios. Our hypothesis is that including invisible watermarks does not always diminish the utility of synthetic medical images and have the potential to be a valuable tool for marking them. Through our research, we aim to emphasize the importance of using watermarks to improve the ethical and legal aspects of medical image synthesis.

\section{Methods}
\label{sec:methods}

\subsection{Data Synthesis Algorithms}
We employed two data synthesis algorithms previously proven effective for data augmentation and pre-training \cite{xing2023you}. The first, StyleGAN, has been a cornerstone in image synthesis within GAN (Generative Adversarial Network) frameworks since 2014. Known for their dual-network adversarial training approach, GAN-based models excel in creating high-quality medical images. For our study, we chose StyleGAN2 for its superior generative modeling in image quality standards. In addition, we introduced Latent Diffusion Models (LDMs) into our study. LDMs perform the diffusion process within a latent space, reducing computational requirements. 

\subsection{The Invisible Watermarking}
In our study, we employed a traditional hybrid watermarking method that combines Discrete Wavelet Transform (DWT) and Discrete Cosine Transform (DCT). We chose to used this frequency based method because of its robustness compared to more recent deep learning-based watermarking methods \cite{iwm-py,zhang2019robust}. In our experiments, the word "synthetic" was used as watermark information and was embedded into a binary sequence.


\subsection{Image Evaluation}

The qualities of images were evaluated using a Python toolbox named MedSynAnalyzer (\url{https://github.com/ayanglab/MedSynAnalyzer}). The fidelity is measured by the ratio of realistic synthetic images to all synthetic images. The variety is measured by the average lossless JPEG size of synthesized images. An illustration of our methods for computing fidelity, variety and privacy is shown in Fig. \ref{fig:utility}. We computed these quality metrics of both watermarked and unwatermarked images. In addition, we also computed the FID score as an additional metrics for analysis. 

To effectively simulate real-world application scenarios of synthetic medical images, we divided our X-ray dataset into four categories: training sets A1 and A2, validation set B, and testing set C. Additionally, we incorporated an open-access pediatric X-ray dataset, D. Within our simulation framework, A1 was treated as a local dataset, while A2 was considered a remote dataset inaccessible to A1. We assessed the value of synthetic data under two distinct conditions:
\begin{enumerate}
    \item  Comparing A1 with synthetic data derived from A1. Here, our objective is to increase the performance of classification models through internal data augmentation.
\item Contrasting A1 with synthetic data produced from A2. In this scenario, our objective is to increase the performance of classification models through external data augmentation.
\end{enumerate}

We subsequently evaluated the performance of AI models using two distinct datasets as is shown in Fig. \ref{fig:utility}. The initial assessment used the local testing set C, which had a consistent clinical background with the training set. In our experiments, we trained the model to distinguish X ray images with and without pleural effusion. Then we tested the model on a different clinical problem to evaluate the generalizability. In our experiments, we investigated whether a model trained on adult X-ray images could adeptly identify key characteristics crucial for the classification of pediatric pneumonia. We first applied these pre-trained model to pediatric X-ray images. Subsequently, features were extracted from the penultimate layer of this model. An SVM classifier was then trained on these features, and its accuracy was evaluated on the test dataset. This approach allowed us to explore the model's flexibility and relevance across diverse medical scenarios.

\section{Experiments}
\label{sec:experiment}
\begin{figure}
    \centering
    \includegraphics[width=\linewidth]{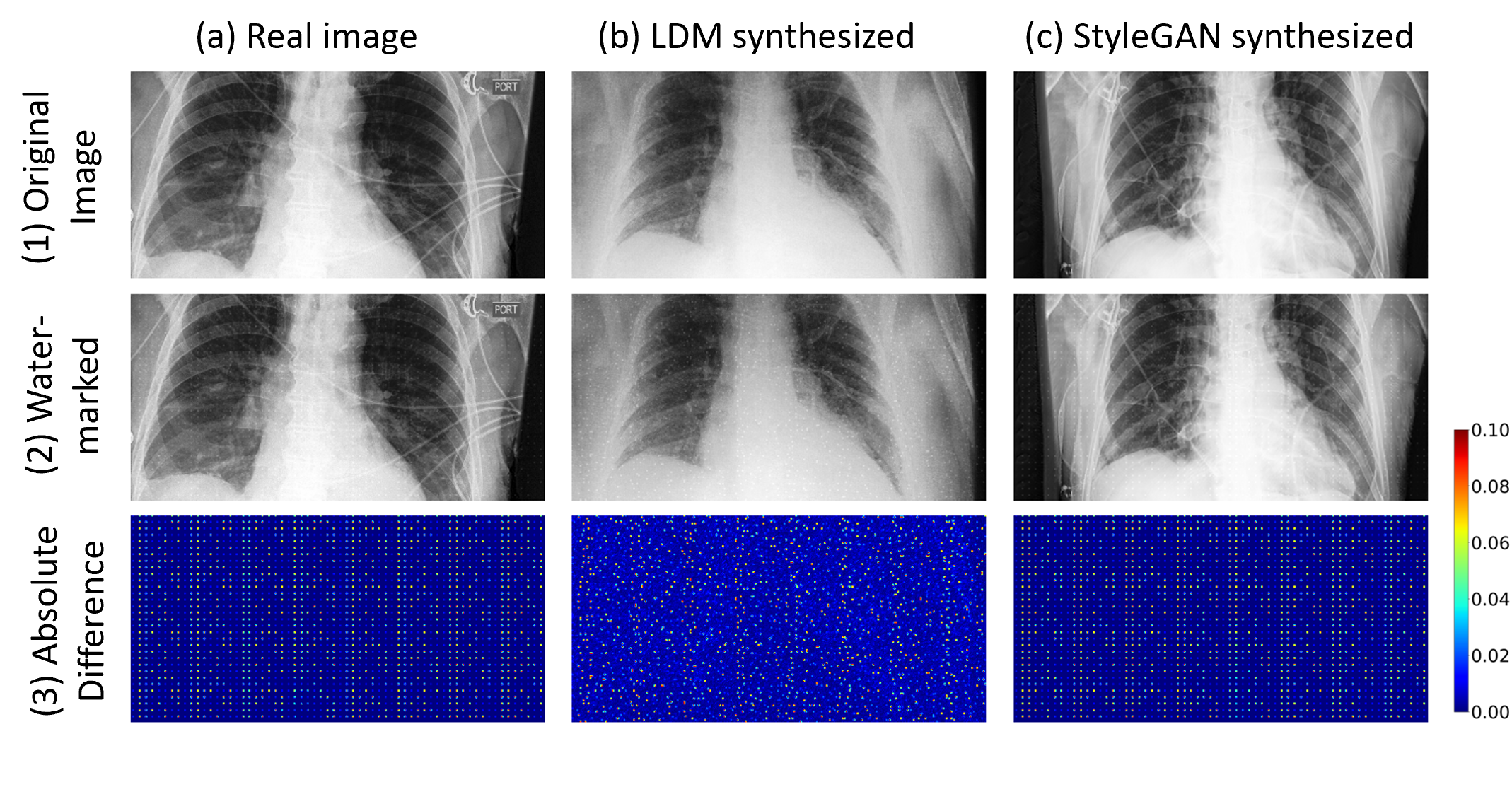}
    \caption{The visualization of watermarks using the DWT+DCT algorithm. All images are normalized to $[0,1]$.}
    \label{fig:vis}
\end{figure}

\subsection{Dataset}
We primarily evaluated the synthetic data's performance with the CheXpert dataset \cite{irvin2019chexpert}, specifically targeting the identification of pleural effusion (PE). For a comprehensive evaluation, we divided this extensive dataset into four subsets: A1 (comprising 15,004 cases with PE and 5,127 without PE), A2 (30,692 with PE and 10,118 without PE), B (3,738 with PE and 1,160 without PE), and C (12,456 with PE and 3,863 without PE). The A1 and A2 sets were partitioned randomly. For both A1 and A2, we generated 20,000 images (10,000 PE and 10,000 without PE) for downstream tasks. 

Additionally, to assess the synthetic models' versatility across different tasks, we employed dataset D, containing 5,863 pediatric X-ray images, encompassing both pneumonia cases and normal controls. All X-ray images were resized to 512×512 pixels and normalized to a range of $[1,1]$.

\subsection{Experimental Setting}
For this study, we utilized two architectures for high resolution medical image synthesis: StyleGAN2 and LDM. In the case of the StyleGAN2 models, synthetic images were generated using a truncation parameter of $\phi=0.6$ to balance between image fidelity and diversity. After synthesis, we trained classification algorithms to evaluate the utility of synthetic images embedded with watermarks. Concurrently, a VQ-VAE model was employed to distill discrete latent features from the images, serving as a feature extractor for quality evaluation. To maintain a fair comparison, we adhered strictly to the parameter configurations as cited in \cite{xing2023you}.

\section{Results and Discussion}
\label{sec:resultsAndDiscussion}
\subsection{Watermark Similarity as an Indicator of Image Fidelity in the Frequency Domain}
In our experiments, we incorporated the term "synthetic" into all synthetic medical images to distinguish them from real ones clearly, as in Fig. \ref{fig:vis}.  We observed that the watermark patterns of real images bear resemblance to those of StyleGAN-generated images. This similarity persists regardless of the differences in watermark information. These observations highlight the frequency likeness between StyleGAN-generated images and real images, substantiating our hypothesis that GAN-based algorithms are capable of producing  images with higher fidelity than LDM. 

\subsection{Watermarking Ensures Uncompromised Synthetic Data Quality}
    
\begin{table}[]
\caption{The quality score of watermarked (WM) images compared to their original copies.}
\label{tab:quality}
\begin{tabular}{p{2cm}cccc}
\hline
Dataset & Fidelity ↑ & Variety ↓ & FID ↓ & Privacy ↑ \\\hline
A2 & 0.345 & 50.949 & 0.266 & / \\
+ WM & 0.348 & 50.993 & 0.266 & / \\\hline
StyleGAN A1 & 0.378 & 166.445 & 15.585 & 0.472 \\
+ WM & 0.377 & 166.369 & 15.584 & 0.472 \\\hline
LDM A1 & 0.091 & 83.736 & 30.117 & 0.985 \\
+ WM & 0.091 & 83.725 & 30.115 & 0.985 \\\hline
StyleGAN A2 & 0.429 & 175.523 & 13.769 & 0.650 \\
+ WM & 0.427 & 175.662 & 13.775 & 0.650 \\\hline
LDM A2 & 0.102 & 84.531 & 27.484 & 0.983 \\
+ WM & 0.102 & 84.538 & 27.482 & 0.983\\\hline
\end{tabular}
\end{table}

In assessing the quality of watermarked synthetic images, we conducted a comprehensive analysis encompassing various metrics, including fidelity, variety, FID score, and the capacity for privacy protection. It was observed that the presence of watermarks only results in negligible alterations in the metric values during evaluation, as in table \ref{tab:quality}. On all datasets, the quality metrics remain consistent after watermarking.

This phenomenon can largely be attributed to the current evaluation algorithms that leverage pre-trained models in an attempt to approximate human perception. These algorithms primarily focus on capturing salient information such as edges, intensity, and large-scale morphology, often overlooking subtle details like watermarks.

\subsection{Watermarking Has an Influence on Synthetic Data Utility}

\begin{figure}
    \centering
    \includegraphics[width=\linewidth]{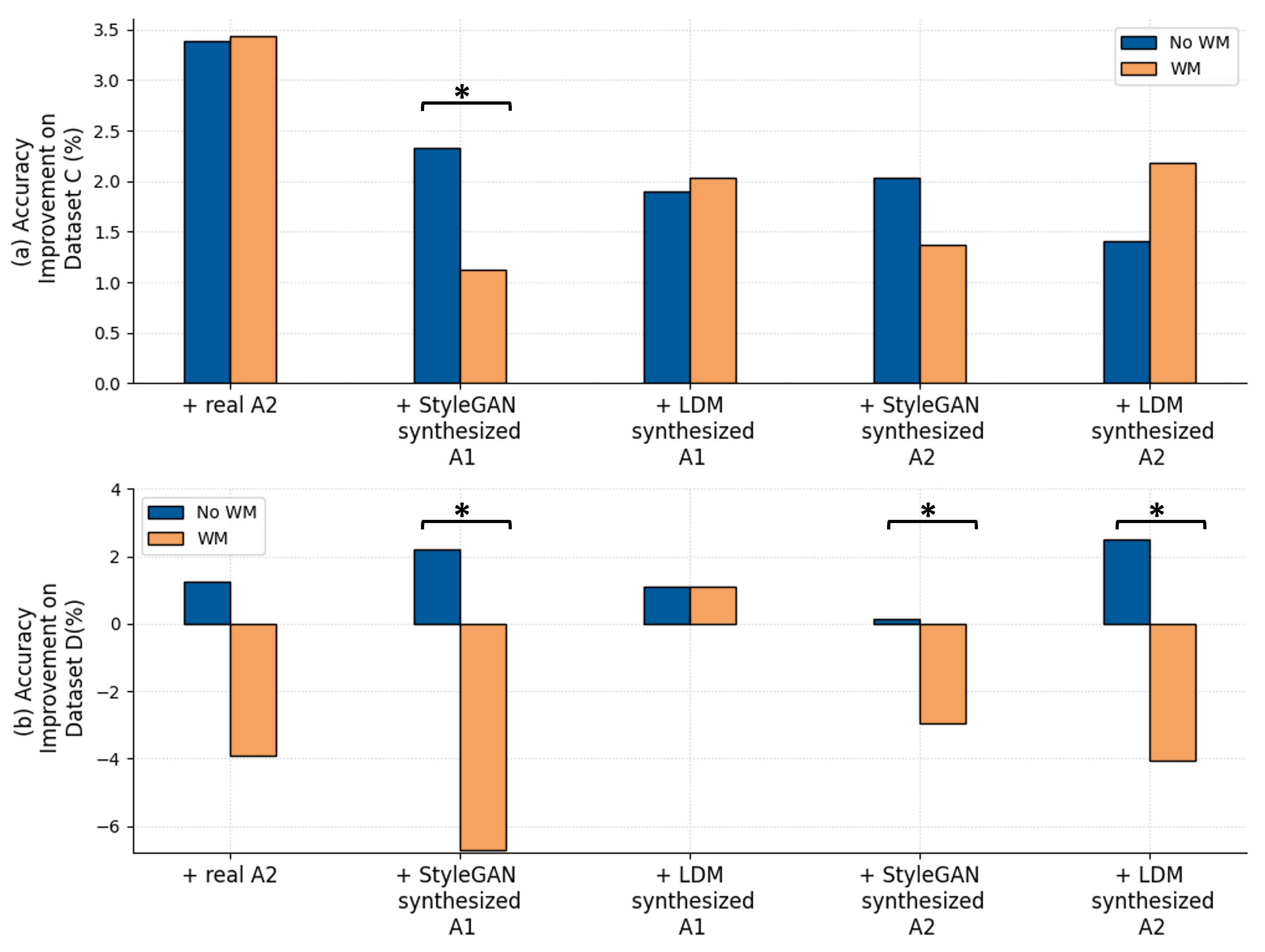}
    \caption{Comparison of accuracy improvement for two evaluation conditions on dataset C (same clinical question) and D (different clinical question). Blue bars represent results when using internal augmentation (StyleGAN synthesized A1 and LDM synthesized A1), and red bars represent results with external augmentation (StyleGAN synthesized A2 and LDM synthesized A2). * represents $p<0.05$ according to McNemar's test. The y-axis is zero-normalized based on the test accuracy of the model trained solely on A1 to provide a clearer comparison.}
    \vspace{-5mm}
    \label{fig:utilitybar}
\end{figure}
As in Section 2.3, we assessed the utility of synthetic images for two objectives: intra-class evaluation and cross-class evaluation. The former evaluates the data augmentation capability, while the latter assesses the generalizability of synthetic data to different clinical questions. For each evaluation type, two scenarios were considered. We compared the performance of a model trained solely on A1 with that of a model trained on A1 supplemented by synthetic data from various sources.

While watermarked images generally maintain the intra-class utility of deep learning algorithms, their generalizability is limited. A significant drop in accuracy was observed when the model trained on watermarked images was tested on dataset under a different clinical background. While watermarked images may effectively represent certain characteristics or patterns relevant to the specific class they belong to, due to the frequency domain manipulation, they may not encapsulate broader features essential for cross-class generalization. 

\section{Conclusion}
In conclusion, our algorithm demonstrates that invisible watermarking can maintain the quality of synthetic data, offering the potential for integration with various data synthesis algorithms to ensure detectability. Our tests indicate that watermarked images retain comparable data augmentation capabilities for intra-class validation similar to their unwatermarked counterparts. However, the limitations of spatial domain quality measurement metrics fail to detect frequency modifications, suggesting these quality metrics may fail to predict the utility of synthetic images. Additionally, when considering outer-class evaluation, the generalizability of models trained using watermarked images could be compromised. Overall, we aim to initiate a deeper discussion regarding the feasibility of adding invisible watermarks to synthetic images. This study can serve as a foundational stepping stone for future investigations in this domain.



\section{Compliance with ethical standards}
This research study was conducted retrospectively using human subject data made available in open access. Ethical approval was not required as confirmed by the license attached with the open access data.
\label{sec:compliance}

\section{acknowledgments}
\label{sec:acknowledgments}
This study was supported in part by the ERC IMI (101005122), the H2020 (952172), the MRC (MC/PC/21013), the Royal Society (IEC/NSFC/211235), the Imperial College Undergraduate Research Opportunities Programme (UROP), the NVIDIA Academic Hardware Grant Program, the SABER project supported by Boehringer Ingelheim Ltd, NIHR Imperial Biomedical Research Centre (RDA01), and the UKRI Future Leaders Fellowship (MR/V023799/1).

\bibliographystyle{IEEEbib}
\bibliography{refs}

\begin{thebibliography}{10}

\bibitem{chawla2002smote}
Nitesh~V Chawla, Kevin~W Bowyer, Lawrence~O Hall, and W~Philip Kegelmeyer,
\newblock ``Smote: synthetic minority over-sampling technique,''
\newblock {\em Journal of artificial intelligence research}, vol. 16, pp.
  321--357, 2002.

\bibitem{coyner2022synthetic}
Aaron~S Coyner, Jimmy~S Chen, Ken Chang, Praveer Singh, Susan Ostmo, RV~Paul
  Chan, Michael~F Chiang, Jayashree Kalpathy-Cramer, J~Peter Campbell, Imaging,
  Informatics in~Retinopathy~of Prematurity~Consortium, et~al.,
\newblock ``Synthetic medical images for robust, privacy-preserving training of
  artificial intelligence: application to retinopathy of prematurity
  diagnosis,''
\newblock {\em Ophthalmology Science}, vol. 2, no. 2, pp. 100126, 2022.

\bibitem{qi2020emerging}
Chang Qi, Jian Zhang, and Peng Luo,
\newblock ``Emerging concern of scientific fraud: Deep learning and image
  manipulation,''
\newblock {\em bioRxiv}, pp. 2020--11, 2020.

\bibitem{hataya2023will}
Ryuichiro Hataya, Han Bao, and Hiromi Arai,
\newblock ``Will large-scale generative models corrupt future datasets?,''
\newblock in {\em Proceedings of the IEEE/CVF International Conference on
  Computer Vision}, 2023, pp. 20555--20565.

\bibitem{fang2016reversible}
Yingying Fang and Bo~Ou,
\newblock ``Reversible data hiding using non-local means prediction,''
\newblock in {\em Algorithms and Architectures for Parallel Processing: ICA3PP
  2016 Collocated Workshops: SCDT, TAPEMS, BigTrust, UCER, DLMCS, Granada,
  Spain, December 14-16, 2016, Proceedings}. Springer, 2016, pp. 125--135.

\bibitem{synthid}
Pushmeet~Kohli Sven~Gowal,
\newblock ``Identifying ai-generated images with synthid,'' 2023,
\newblock
  \url{https://www.deepmind.com/blog/identifying-ai-generated-images-with-
  synthid} [Accessed: (Oct. 24th, 2023)].

\bibitem{saharia2022photorealistic}
Chitwan Saharia, William Chan, Saurabh Saxena, Lala Li, Jay Whang, Emily~L
  Denton, Kamyar Ghasemipour, Raphael Gontijo~Lopes, Burcu Karagol~Ayan, Tim
  Salimans, et~al.,
\newblock ``Photorealistic text-to-image diffusion models with deep language
  understanding,''
\newblock {\em Advances in Neural Information Processing Systems}, vol. 35, pp.
  36479--36494, 2022.

\bibitem{fernandez2023stable}
Pierre Fernandez, Guillaume Couairon, Herv{\'e} J{\'e}gou, Matthijs Douze, and
  Teddy Furon,
\newblock ``The stable signature: Rooting watermarks in latent diffusion
  models,''
\newblock {\em arXiv preprint arXiv:2303.15435}, 2023.

\bibitem{xing2023you}
Xiaodan Xing, Federico Felder, Yang Nan, Giorgos Papanastasiou, Walsh Simon,
  and Guang Yang,
\newblock ``You don't have to be perfect to be amazing: Unveil the utility of
  synthetic images,''
\newblock {\em arXiv preprint arXiv:2305.18337}, 2023.

\bibitem{iwm-py}
Qingquan Wang,
\newblock ``invisible-watermark 0.2.0,'' 2023,
\newblock \url{https://pypi.org/project/invisible-watermark/} [Accessed: (Oct.
  24th, 2023)].

\bibitem{zhang2019robust}
Kevin~Alex Zhang, Lei Xu, Alfredo Cuesta-Infante, and Kalyan Veeramachaneni,
\newblock ``Robust invisible video watermarking with attention,''
\newblock 2019.

\bibitem{irvin2019chexpert}
Jeremy Irvin, Pranav Rajpurkar, Michael Ko, Yifan Yu, Silviana Ciurea-Ilcus,
  Chris Chute, Henrik Marklund, Behzad Haghgoo, Robyn Ball, Katie Shpanskaya,
  et~al.,
\newblock ``Chexpert: A large chest radiograph dataset with uncertainty labels
  and expert comparison,''
\newblock in {\em Proceedings of the AAAI conference on artificial
  intelligence}, 2019, vol.~33, pp. 590--597.

\end{thebibliography}

\end{document}